\documentclass{emulateapj}%
\shorttitle{Early UV Ingress in WASP-12b: Measuring Planetary Magnetic Fields}
\shortauthors{Vidotto, Jardine \& Helling}
\defcitealias{lai2010}{LHV10}
\defcitealias{hebb2009}{H09}

\begin{document}
\title{Early UV Ingress in WASP-12b: Measuring Planetary Magnetic Fields}
\author{A.~A.~Vidotto} 
\affil{School of Physics and Astronomy, University of St Andrews, North Haugh, St Andrews, KY16 9SS, UK}
\email{Aline.Vidotto@st-andrews.ac.uk} 
\and
\author{M.~Jardine} 
\affil{School of Physics and Astronomy, University of St Andrews, North Haugh, St Andrews, KY16 9SS, UK}
\and 
\author{Ch.~Helling}
\affil{School of Physics and Astronomy, University of St Andrews, North Haugh, St Andrews, KY16 9SS, UK}

\begin{abstract}
Recently, \citeauthor{fossati2010} observed that the UV transit of WASP-12b showed an early ingress compared to the optical transit. We suggest that the resulting early ingress is caused by a bow shock ahead of the planetary orbital motion. In this Letter we investigate the conditions that might lead to the formation of such a bow shock. We consider two scenarios: (1) the stellar magnetic field is strong enough to confine the hot coronal plasma out to the planetary orbit and (2) the stellar magnetic field is unable to confine the plasma, which escapes in a wind. In both cases, a shock capable of compressing plasma to the observed densities will form around the planet for plasma temperatures $T \lesssim (4 - 5) \times 10^6~$K. In the confined case, the shock always forms directly ahead of the planet, but in the wind case the shock orientation depends on the wind speed and hence on the plasma temperature. For higher wind temperatures, the shock forms closer to the line of centers between the planet and the star. We conclude that shock formation leading to an observable early UV ingress is likely to be a common feature of transiting systems and may prove to be a useful tool in setting limits on planetary magnetic field strengths $B_p$. In the case of WASP-12b, we derive an upper limit of about $B_p=24$~G. 
\end{abstract}

\keywords{planet-star interactions --- planets and satellites: individual (WASP-12b) --- planets and satellites: magnetic fields --- stars: coronae --- stars: individual (WASP-12) --- stars: winds, outflows}

\section{Introduction}
WASP-12b is among the largest transiting planets discovered so far. First identified in an optical photometric transit survey \citep[][hereafter, H09]{hebb2009}, its mass is $M_p=1.41~M_J$ and radius $R_p = 1.79~R_J$, where $M_J$ and $R_J$ are the mass and radius of Jupiter, respectively. After its discovery, additional observations have been acquired \citep[e.g., ][]{fossati2010, fossati2010b, lopes2010, 2010arXiv1004.1809H, 2010arXiv1003.2763C}. Interestingly, transit observations in the near-UV revealed a longer transit duration than in the optical \citep{fossati2010}. While the time of the egress occurs almost simultaneously for both set of observations, the ingress of the transit is first seen in the near-UV wavelength range. This asymmetric behavior has been explained by the presence of asymmetries in the planetary atmosphere.

Close-in giant gas planets are rather inflated and most have developed an exosphere (HD~209458b: \citealt{vidal2003,vidal2008, ehrenreich2008, linsky2010}; HD~189733b: \citealt{lecavelier2010}; CoRoT-Exo-1b: \citealt{barge2008}; WASP-12b: \citealt{fossati2010}) that can fill or even overflow the planet's Roche lobe \citep{gu2003,li2010,Ibgui}. This may result in mass transfer through a Lagrangian point to the star that could cause an asymmetry in the appearance of the transiting planet-star system as seen from the Earth \citep[][hereafter, LHV10]{lai2010}. Asymmetries could also be produced by cometary tails. However, \citet{ehrenreich2008} demonstrated for HD~209458b that a radiation-driven cometary tail would produce a late egress of the planetary transit light curve, instead of an early ingress. \citetalias{lai2010} investigated the formation of a bow shock around the planet due to the interaction of the planet's magnetosphere with a stellar wind as the cause of the early ingress. Assuming a typical solar wind mass-loss rate and adopting solar wind properties, \citetalias{lai2010} derived the wind velocity at the distance of WASP-12b using a thermally driven wind model \citep{parker}. They found that the wind is still subsonic at WASP-12b orbital distance, concluding that no bow shock is expected to form in the planet-wind interaction zone. Here, we demonstrate that a bow shock can actually be formed around the planet if the relative azimuthal velocity between the planetary orbital motion and the ambient medium is taken into account.

WASP-12b orbits its host star (a late-F main-sequence star, with mass $M_* = 1.35~M_\odot$ and radius $R_*=1.57~R_\odot$) at an orbital radius of $R_{\rm orb}=0.023~{\rm AU}=3.15~R_*$, with an orbital period of $P_{\rm orb} = 1.09$~d \citepalias{hebb2009}. Due to its close proximity to the star, the flux of coronal particles impacting on the planet comes mainly from the azimuthal direction, as the planet moves at a Keplerian orbital velocity of $u_K = (G M_*/R_{\rm orb})^{1/2}\sim 230$~km~s$^{-1}$ around the star. Therefore, stellar coronal material is compressed ahead of the planetary orbital motion, possibly forming a bow shock ahead of the planet. If such compressed material is optically thin, the planetary transit light curve is symmetrical with respect to the ingress and the egress of the planet. Indeed, this is the case when the transit is observed at optical wavelengths \citepalias{hebb2009}. However, if the shocked material ahead of the planet can absorb enough stellar radiation, the observer will note an early ingress of the planet in the stellar disk, but no difference will be seen at the time of egress, as the shocked material is present only ahead of the planetary motion. 

Here, we investigate under which conditions the interaction of a planet with the stellar coronal plasma could lead to the formation of a bow shock ahead of the planetary orbital motion, and therefore, explain the early ingress observable in the near-UV. Our shock model is described in Section~\ref{sec.shockbasics}. Although we know the orbital radius of the planet, we do not know if at this radius the stellar magnetic field is still capable of confining the hot gas of the stellar corona, or if this gas is escaping in a wind. Therefore, we investigate the validity of our model with respect to different stellar coronal conditions in Sections~\ref{sec.corona} and \ref{sec.wind}. Conclusions are presented in Section~\ref{sec.conclusions}.

\section{The Shock Model}\label{sec.shockbasics}
\begin{figure*} 
\includegraphics[width=180mm]{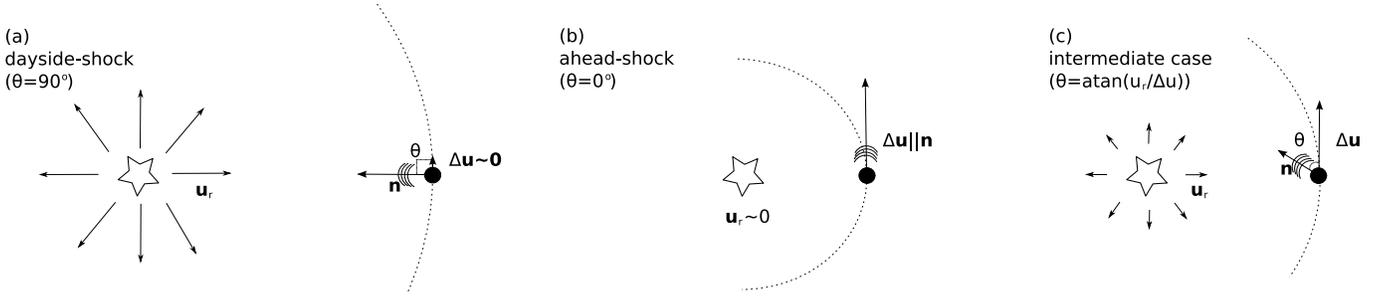}	
\caption{Sketch of shock types (not to scale): (a) dayside-shock ($\theta=90^{\rm o}$), (b) ahead-shock ($\theta=0^{\rm o}$), and (c) intermediate case. Arrows radially leaving the star depict the stellar wind, dashed semi-circles represent the orbital path, $\theta$ is the deflection angle between ${\bf n}=\Delta {\bf u} - {\bf u}_r$ and the relative azimuthal velocity of the planet $\Delta {\bf u}$. \label{fig.shockconditions} }
\end{figure*}

A bow shock around a planet is formed when the relative motion between the planet and the stellar corona/wind is supersonic. The shock configuration depends on the direction of the flux of particles that arrives at the planet. We illustrate two different limits of the shock configuration and an intermediate case in Figure~\ref{fig.shockconditions}, where $\theta$ is the deflection angle between the azimuthal direction of the planetary motion and ${\bf n}$, a vector that defines the outward direction of the shock. As seen from the planet, $-{\bf n}$ is the velocity of the impacting material. 

The first shock limit, a ``dayside-shock'', occurs when the dominant flux of particles impacting on the planet arises from the (radial) wind of its host star. For instance, the impact of the supersonic solar wind forms a bow shock at the dayside of Earth's magnetosphere (i.e., at the side that faces the Sun). This condition is illustrated in Figure~\ref{fig.shockconditions}(a) and is met when $u_r > c_s$, where $u_r$ and $c_s$ are the local radial stellar wind velocity and sound speed, respectively.

A second shock limit, an ``ahead-shock'', occurs when the dominant flux of particles impacting on the planet arises from the relative azimuthal velocity between the planetary orbital motion and the ambient plasma. This condition is especially important when the planet orbits at a close distance to the star, and therefore, possesses a high $u_K$. In this case, the velocity of the particles that the planet `sees' is supersonic if $\Delta u = |u_K-u_\varphi|> c_s$, where $u_\varphi$ is the azimuthal velocity of the stellar corona. This condition is illustrated in Figure~\ref{fig.shockconditions}(b).

For intermediate cases, both the wind and the azimuthal relative velocities will contribute to the formation of a shock around the planet (Figure~\ref{fig.shockconditions}(c)) and the deflection angle $\theta$ is given by 
\begin{equation}  \label{eq.theta}
\theta = {\rm atan}{ \left( \frac{u_r}{|u_K-u_\varphi|} \right)} .
\end{equation} 
In general, for a planet orbiting its host star at a close distance, the stellar wind is still accelerating and subsonic ($u_r<c_s$). In this case, conditions for an ahead-shock will more probably be met.

Here, we adopt the estimates of \citetalias{lai2010} to derive the minimum density of the stellar plasma at the orbital radius of WASP-12b. From the observations presented in \citet{fossati2010}, \citetalias{lai2010} estimated the column density of the absorbing gas around the planet to be $ \gtrsim 1.4 \times 10^{13}~\mbox{cm}^{-2}$. For that, they assumed $\tau=1$ in the absorption lines of Mg. Because the time resolution of the transit measurement is sparse, it is difficult to place a firm constrain on the thickness of the absorbing material, which could either extend all the way to the planetary surface or be limited to a geometrically thin shocked region. Through time differences of the optical/near-UV ingresses, \citetalias{lai2010} estimated that the stand-off distance between the shock and the center of the planet is about $4.2~R_p$, leading to a Mg density $n_{\rm Mg} \gtrsim 400~\mbox{cm}^{-3}$. Here, we convert this minimum density to hydrogen number density by using the observed metallicity of WASP-12, $[{\rm M}/{\rm H}]=0.3$ \citepalias{hebb2009}
\begin{equation}
\frac{n_{\rm Mg}}{n_H} =10^{(\epsilon_{\rm Mg} - \epsilon_H)} 10^{0.3 } \simeq 6.76 \times 10^{-5} ,
\end{equation}
where $\epsilon_{\rm Mg} = 7.53$ and $\epsilon_H  = 12.00$ are the solar element abundances for Mg and H, respectively \citep{grevesse2007}. Therefore, the minimum density of the shocked stellar coronal plasma is 
\begin{equation}\label{eq.shock-density}
{n_H} \simeq \frac{400~{\rm cm}^{-3}}{6.76 \times 10^{-5}} \simeq 6 \times 10^{6}~{\rm cm}^{-3} .
\end{equation}
The density behind such a shock would, in the adiabatic limit, be at most four times the density ahead of the shock. From Equation~(\ref{eq.shock-density}), this would then require a pre-shock coronal density of
\begin{equation}\label{eq.coronal-density}
{n_{\rm obs}} \simeq 1.5 \times 10^{6}~{\rm cm}^{-3} 
\end{equation}
in the external ambient medium at the orbital radius of the planet.

The temperature of the absorbing material must be such as to allow for the presence of Mg II lines. For an adiabatic shock, the temperature immediately behind the shock depends on the squared Mach number of the flow impacting on the planet. To determine the temperature and ionization structures of the atmosphere of the planet, and therefore quantitatively evaluate the optical depth of Mg II lines, detailed radiative transfer models are needed.

\section{Hydrostatic Corona}\label{sec.corona}
\begin{figure}
\includegraphics[width=84mm]{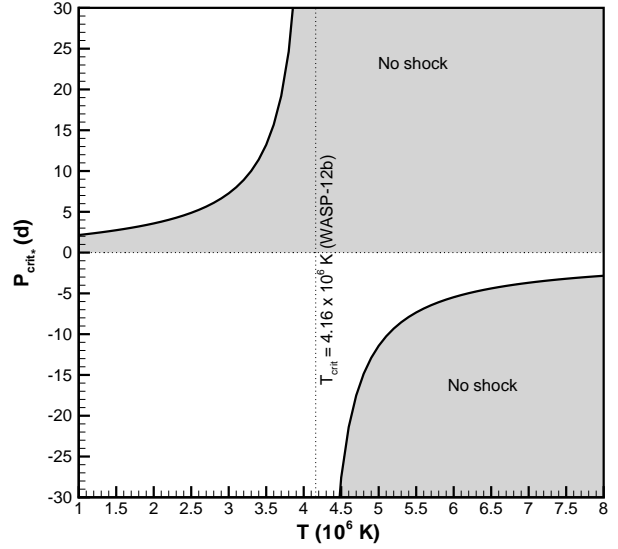}
\caption{Critical stellar rotation period (Equation~(\ref{eq.pcrit})) required for the formation of a bow shock as a function of the environment temperature. Negative periods refer to cases where the planet is in a retrograde orbit.}
\label{min_prot_T} 
\end{figure}

In this section, we assume that the orbit of the planet lies within the stellar magnetosphere. In this case, the planet moves through a medium that is confined by a rigid stellar magnetic field and so rotates at the stellar rotation rate. The relative azimuthal velocity between the planet and this medium is therefore 
\begin{equation}\label{eq.deltau}
\Delta u = |u_K - u_{\varphi,{\rm cor}}| = \left| \left( \frac{G M_*}{R_{\rm orb}} \right)^{1/2} - \frac{2 \pi R_{\rm orb}}{P_*}\right| ,
\end{equation}
where $u_{\varphi,{\rm cor}} = 2 \pi R_{\rm orb}/P_*$ is the velocity of the medium corotating with the star, and $P_*$ is the stellar period of rotation. If $\Delta u> c_s$, a bow shock forms around the planet. We may therefore define a critical stellar rotation period such that $\Delta u = c_s$ and
\begin{equation}\label{eq.pcrit}
|P_{{\rm crit},*}| =  \frac{2\pi R_{\rm orb}}{\left|u_K- c_s \right| } .
\end{equation}
Figure~\ref{min_prot_T} presents $P_{{\rm crit},*}$ as a function of the coronal temperature $T$. The shaded areas show regions on the $P_{{\rm crit},*} - T$ parameter space where no shock is formed. In our notation, $P_{{\rm crit},*}<0$ refers to cases where the star is counterrotating with respect to the planetary orbital rotation \citep[e.g., WASP-8b;][]{queloz2010}. The vertical line in Figure~\ref{min_prot_T} represents a critical coronal temperature where $u_K=c_s$ and $|P_{{\rm crit},*}| \to \infty$ (cf. Equation~(\ref{eq.pcrit})). For WASP-12, we obtain
\begin{equation}\label{eq.tcrit}
T_{{\rm crit}} =  \frac{G M_* m}{R_{\rm orb} k_B} = 23.1\times 10^6 \frac{(M_*/M_\odot) \mu }{(R_{\rm orb}/R_\odot)} ~{\rm K}= 4.16 \times 10^6~{\rm K} ,
\end{equation}
where $m=\mu m_p$ is the mean particle mass, $k_B$ is the Boltzmann constant, and $m_p$ is the proton mass. We adopt $\mu=0.66$. Above this temperature, no shock formation is possible at any positive $P_*$. For all the remaining areas of the plot an ahead-shock will develop.\footnote{A ``behind-shock'' is formed when $u_K - u_{\varphi,{\rm cor}}<0$, or $P_{{\rm crit},*}<P_{\rm orb}\simeq 1$~d for WASP-12b. In such situation, the planet orbits beyond the Keplerian corotation radius. The coronal plasma lags behind the planetary motion and the shock trails the planet. Observationally, we would detect a late egress, instead of an early ingress. However, Figure~\ref{min_prot_T} shows that for $P_{{\rm crit},*}<1$~d (WASP-12b), the relative azimuthal velocity $\Delta u$ is subsonic, therefore, no shock will be formed.} Although $P_*$ is unknown, WASP-12 is likely to be a slow rotator \citep{fossati2010b}. With this information, the presence of an ahead-shock for WASP-12b constrains the coronal temperature to be $T \lesssim (4 - 5) \times 10^6$~K.

\begin{figure}
	\includegraphics[width=84mm]{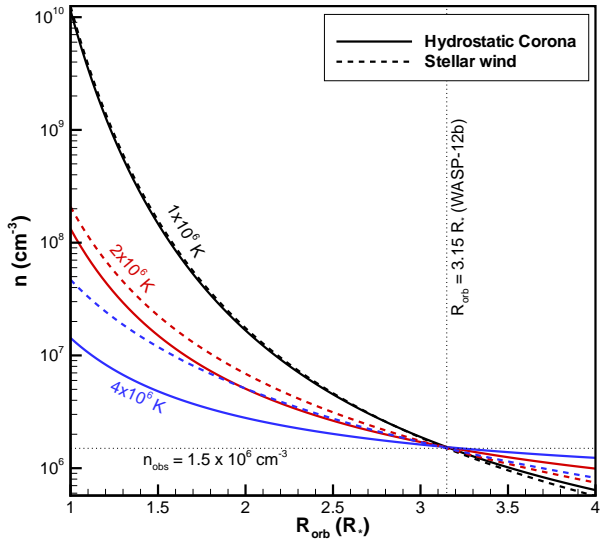}
	\caption{Coronal density of an isothermal plasma for three different temperatures. \label{fig.wind-density}}
\end{figure}

For an isothermal corona, the density is 
\begin{eqnarray}\label{eq.dens-hyd}
 n(R_{\rm orb}) = n_0\exp\left[ \frac{GM_*/R_*}{k_B T/m} \left( \frac{R_*}{R_{\rm orb}} -1 \right) \right. \nonumber \\ 
\left . + \frac{2 \pi^2R_*^2/P_*^2}{k_B T/m} \left( \frac{R_{\rm orb}^2}{R_*^2}-1 \right)\right] ,
\end{eqnarray}
where $n_0$ is the density at the coronal base. From Equation~(\ref{eq.dens-hyd}), we note that the more slowly the star rotates, the lower is the density at any given radius. For the very slow stellar rotation rates expected for stars with detected transiting planets, $P_*$ has little influence on the density. The solid lines in Figure~\ref{fig.wind-density} show the density variation for a range of orbital radius for three different temperatures and a solar value $P_*=26$~d. The density was scaled to match the observationally derived value of ${n_{\rm obs}} \simeq 1.5 \times 10^{6}~{\rm cm}^{-3}$ (Equation~(\ref{eq.coronal-density})), which is the minimum density required for the detection of the early ingress. Except for the very low temperature case (black solid line), a typical solar coronal base density \citep[i.e., $n_0 \sim 10^8~{\rm cm}^{-3}$; ][]{withbroe1988} provides the condition for the existence of a shock. 

We note that the observationally derived stand-off distance from the shock to the center of the planet can be taken as approximately the extent of the planetary magnetosphere $r_M\simeq 4.2~R_p$ \citepalias{lai2010}. Pressure balance between the coronal total pressure and the planet total pressure requires that, at $r_M$,
\begin{equation}\label{eq.equilibrium}
\frac{\rho_c \Delta u^2}{2} + \frac{B_c(R_{\rm orb})^2}{4\pi} + p_c= \frac{B_{p}(r_M)^2}{4\pi} + p_{p} ,
\end{equation}
where $\rho_c=m n_{\rm obs}$, $p_c$ and $B_c(R_{\rm orb})$ are the local coronal mass density, thermal pressure, and magnetic field intensity, and $p_p$ and $B_p (r_M)$ are the planet thermal pressure and magnetic field intensity at $r_M$. Neglecting the kinetic term and the thermal pressures\footnote{It is straightforward, however, to show that the kinetic term and both thermal pressure terms are negligible relative to the magnetic pressure terms. From the estimated planetary magnetic field ($24$~G), the magnetic pressure is $p_B \simeq {B_{p}(r_M)^2}/({4\pi}) \simeq 8 \times 10^{-3}$~dyn~cm$^{-2}$. The kinetic term ${\rho_c \Delta u^2}/{2}$ in Equation~(\ref{eq.equilibrium}) is dependent on the period of the star, which for WASP-12 is unknown. For example, for $P_*=26$~d, ${\rho_c \Delta u^2}/{2} \simeq 4\times 10^{-4}$~dyn~cm$^{-2}$. For $T=2\times 10^6$~K, $p_c \simeq 4\times 10^{-4}$~dyn~cm$^{-2}$. To estimate the thermal pressure of the planet, we use values derived from \citet{murray-clay2009} for planetary density ($\sim 10^7$~cm$^{-3}$) and temperature ($10^{3.5}$~K), which result in $p_p \simeq 4\times 10^{-6}$~dyn~cm$^{-2}$. Because $p_B$ is much larger than the aforementioned pressures, the magnetic terms in Equation~(\ref{eq.equilibrium}) will dominate.}, we may therefore rewrite previous equation as
\begin{equation}\label{eq.equilibrium1}
B_c(R_{\rm orb}) \simeq B_p(r_M) ,
\end{equation}
from which we can estimate an upper limit for the planetary magnetic field intensity.

For a stellar dipolar magnetic field, ${B_c(R_{\rm orb})}= B_* \left( {R_*}/{R_{\rm orb}}\right)^3$. Assuming that the planetary magnetic field is also dipolar, then ${B_p (r_M)}= B_p \left( {R_p}/{r_M}\right)^3$. $B_*$ and $B_p$ are the magnetic field intensities at the stellar and planetary surfaces, respectively. From Equation~(\ref{eq.equilibrium1}), we have
\begin{equation}
B_p = B_* \left(  \frac{R_*/R_{\rm orb}}{R_{p}/r_M}  \right)^3 = B_* \left(  \frac{1/3.15}{1/4.2}  \right)^3\simeq 2.4 B_* ,
\end{equation}
where we have used the observationally derived characteristics for the WASP-12 system: $r_M=4.2~R_p$ and $R_{\rm orb}= 3.15~R_*$. So far, tentative measurements of the stellar magnetic field have not provided significant detections \citep[for theoretical predictions, see][]{christensen2009}, but have yielded an upper limit of $10$~G in the longitudinal component \citep{fossati2010b}. Adopting this upper limit as $B_*$, our model predicts a maximum planetary magnetic field of about $24$~G. 
 
\section{Stellar Wind}\label{sec.wind}
\begin{figure}
	\includegraphics[width=84mm]{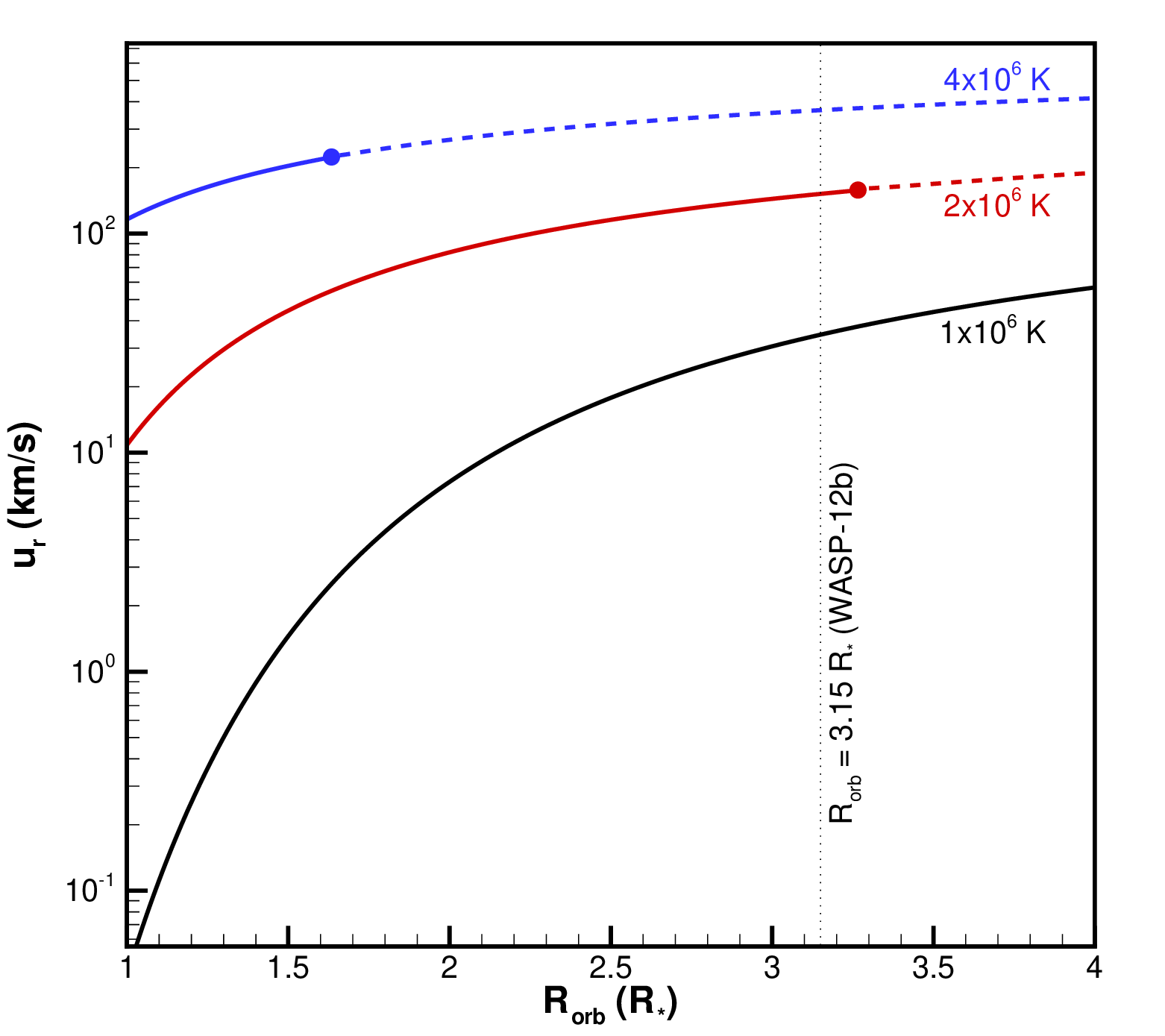}
	\caption{Coronal wind velocity of an isothermal wind for three different temperatures. Filled circles represent the location of the sonic point ($u_r=c_s$) and dashed lines represent the regions of supersonic velocities, which, for $T=1\times 10^6~$K, is outside the range of orbital radii shown above. \label{fig.wind-velocity} }
\end{figure}

In this section, we investigate the scenario where the stellar magnetic field is not strong enough to confine the coronal plasma, which expands in the form of stellar wind. For a hot corona, the thermal pressure gradient is able to drive a wind \citep{parker}. We consider the simplest case of a thermally-driven wind which is not magnetically channeled. The wind radial velocity $u_r$ is derived from the integration of the differential equation
\begin{equation}  
\rho u_r \frac{\partial u_r}{\partial r} = -\frac{\partial p}{\partial r} - \rho \frac{G M_*}{r^2} . 
\end{equation} 
Figure~\ref{fig.wind-velocity} presents the wind velocity profile at different orbital radius up to a few stellar radii for three isothermal wind temperatures. In the distance range $R_{\rm orb}<4R_*$, the wind is still accelerating, and the radial velocities are considerably smaller than the terminal velocities $u_\infty$ achieved: $u_\infty \simeq 440$, $670$ and $1020$~km~s$^{-1}$ for $T=1$, $2$, and $4\times10^6~$K, respectively. The filled circles are the sound speed of the wind: $c_s \simeq 158$ and $223$~km~s$^{-1}$ for $T=2$ and $4\times10^6~$K. Dashed lines represent the regions of supersonic velocities. For the lowest-temperature case ($1\times10^6~$K), the wind becomes supersonic at a much larger radii, not shown in Figure~\ref{fig.wind-velocity}. The vertical dotted line shown in Figure~\ref{fig.wind-velocity} represents the orbital radius of WASP-12b. 

The density structure of the wind is derived from the conservation of mass, where $n u_r r^2$ is a constant of the steady-state wind. The dashed lines in Figure~\ref{fig.wind-density} present the density profile for an isothermal wind, for three coronal temperatures. In contrast to the hydrostatic corona (solid lines), $n$ does not depend on $P_*$. Except for the lowest temperature adopted, where the density is well described by a hydrostatic density profile, the coronal base density $n_0$ required for the formation of a shock in the wind case is larger than the one required by the hydrostatic corona case. Nevertheless, as in the static scenario, a typical solar coronal density still allows for the formation of a shock.

In the stellar wind case, there are no azimuthal forces acting on the flow. Therefore, through conservation of angular momentum of the particles leaving the star ($2 \pi R_*^2/P_*$), the azimuthal velocity of the wind is $u_{\varphi, {\rm wind}} = 2 \pi R_*^2 /(P_* R_{\rm orb})$. This implies that $\Delta u$ for the wind case is 
\begin{equation}
\Delta u = |u_K -u_{\varphi, {\rm wind}}| = \left| \left( \frac{G M_*}{R_{\rm orb}} \right)^{1/2} - \frac{2 \pi R_*^2 }{P_* R_{\rm orb}}\right| ,
\end{equation}
which is independent of the wind temperature or radial velocity. For slow rotators, $\Delta u$ becomes almost independent of $P_*$ and is given by $\Delta u \approx u_K$.

The condition for shock formation requires that $(u_r^2 + \Delta u^2)^{1/2}> c_s$, implying that
\begin{equation}\label{eq.pcrit2}
|P_{{\rm crit},*}| =  \frac{2\pi R_*^2/R_{\rm orb}}{( u_K^2 + u_r^2 - c_s^2 )^{1/2} } ,
\end{equation}
which, qualitatively, produces a similar result as the one presented for the static case (Figure~\ref{min_prot_T}). This means that the critical temperature derived in the previous section (Equation~(\ref{eq.tcrit})) is enhanced by a temperature $\delta T = u_r^2 m /k_B \simeq  8 \times 10^5 (u_r/(100~{\rm km~s}^{-1}))^2$~K, implying that for $T \lesssim T_{\rm crit} + \delta T$, a shock will be formed.

\begin{figure}
	\includegraphics[width=84mm]{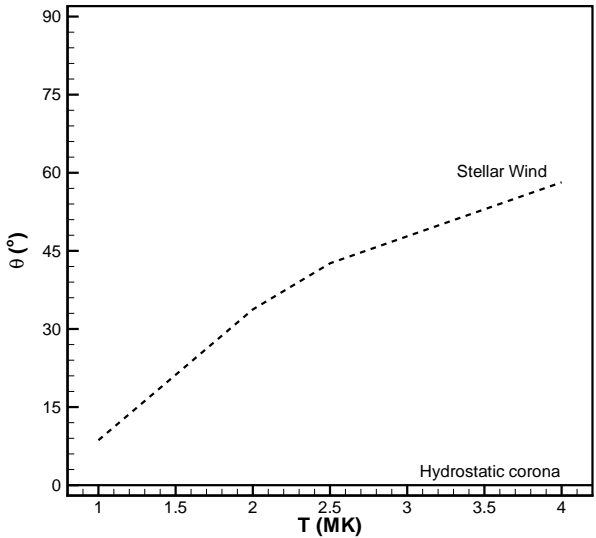}
	\caption{Angle that the shock normal ${\bf n}$ makes to the relative azimuthal velocity of the planet. \label{fig.theta}}
\end{figure}

In the wind case, Figure~\ref{fig.wind-velocity} shows that $u_r \ne 0$ at the orbital radius of WASP-12b, implying that the shock will not form directly ahead of the planet, but at an intermediate angle $\theta$. Because $\theta$ depends on $u_r$ (Equation~(\ref{eq.theta})), which depends on the temperature, the orientation of the shock is temperature-dependent. Figure~\ref{fig.theta} shows how the angle $\theta$ depends on the wind temperature for $P_*=26$~d. While in the hydrostatic case, we found that an ahead-shock ($\theta =0^{\rm o}$) is always formed, the wind case requires a very low temperature to form an ahead-shock. For higher wind temperatures, the shock normal gets closer to the line of centers between the planet and the star and $\theta$ approaches $90^{\rm o}$.

\section{Conclusion}\label{sec.conclusions}
Motivated by the recent observation on the light curve asymmetry of the transit hot-Jupiter WASP-12b \citep{fossati2010}, we proposed a model where the interaction of the stellar plasma with the planet results in the formation of a bow shock around the planet, which could explain the early transit ingress observed in the near-UV. Although we know the orbital radius of the planet, we do not know if at this radius the stellar magnetic field is still capable of confining the hot gas of its corona, or if this gas is escaping in a wind. Therefore, we investigated the physical conditions of the external ambient medium around the planet that could allow for the formation of such a shock. 

In either case, for plasma temperatures  $T \lesssim (4 - 5) \times10^6~$K we expect that a shock capable of compressing the plasma to the observed densities will form around the planet. In the case where the coronal plasma is confined, we showed that the geometry of this shock is independent of the plasma temperature. The shock forms ahead of the planet in its orbital path. In the unconfined (wind) case, the angle between the shock normal and the direction of planetary motion depends on the ratio of the radial wind speed and the azimuthal speed of the planet relative to the stellar wind. Since the wind speed depends on the temperature, the orientation of the shock is temperature-dependent. 

If the planet's orbit takes it through both regions of confined plasma and also wind plasma, the orientation and density of the shock may change,  producing time-dependent absorption and duration of transit. 

Given the large range of stellar rotation rates and coronal/wind temperatures at which a shock is capable of producing the densities estimated by \citetalias{lai2010}, we conclude that this is likely to be a common feature of transiting systems and may prove to be a useful tool in setting limits on planetary magnetic field strengths. For the case of WASP-12b, we derived an upper limit for the planet's magnetic field of about $24$~G. 


\end{document}